\documentclass[
 reprint,superscriptaddress
]{revtex4-2}
\usepackage[colorlinks=true, allcolors=blue]{hyperref}
\usepackage{graphicx}
\usepackage{dcolumn}
\usepackage{bm}

\begin{document}

\preprint{APS/123-QED}

\title{Magneto-optical trapping of aluminum monofluoride}

\author{J. E. Padilla-Castillo}
\affiliation{Fritz-Haber-Institut der Max-Planck-Gesellschaft, Faradayweg 4-6, 14195 Berlin, Germany}
\author{J. Cai}
\affiliation{Fritz-Haber-Institut der Max-Planck-Gesellschaft, Faradayweg 4-6, 14195 Berlin, Germany}
\author{P. Agarwal}
\affiliation{Fritz-Haber-Institut der Max-Planck-Gesellschaft, Faradayweg 4-6, 14195 Berlin, Germany}
\author{P. Kukreja}
\affiliation{Fritz-Haber-Institut der Max-Planck-Gesellschaft, Faradayweg 4-6, 14195 Berlin, Germany}
\author{R. Thomas}
\affiliation{Fritz-Haber-Institut der Max-Planck-Gesellschaft, Faradayweg 4-6, 14195 Berlin, Germany}
\author{B. G. Sartakov}
\affiliation{Fritz-Haber-Institut der Max-Planck-Gesellschaft, Faradayweg 4-6, 14195 Berlin, Germany}
\author{S. Truppe}
\affiliation{Centre for Cold Matter, Blackett Laboratory, Imperial College London, London SW7 2AZ}
\author{G. Meijer}
\affiliation{Fritz-Haber-Institut der Max-Planck-Gesellschaft, Faradayweg 4-6, 14195 Berlin, Germany}
\author{S. C. Wright}
\email[]{sidwright@fhi-berlin.mpg.de}
\affiliation{Fritz-Haber-Institut der Max-Planck-Gesellschaft, Faradayweg 4-6, 14195 Berlin, Germany}

\date{\today}

\begin{abstract}
Magneto-optical trapping of molecules has thus far been restricted to molecules with $^2\Sigma$ electronic ground states. These species are chemically reactive and only support a simple laser cooling scheme from their first excited rotational level. Here, we demonstrate a magneto-optical trap (MOT) of aluminum monofluoride (AlF), a deeply bound and intrinsically stable diatomic molecule with a $^1\Sigma^+$ electronic ground state. The MOT operates on the strong A$^1\Pi\leftarrow{}$X$^1\Sigma^+$ transition near 227.5~nm, whose Q$(J)$ lines are all rotationally closed. We demonstrate a MOT of about $6\times 10^4$ molecules for the $J=1$ level of AlF, more than $10^4$ molecules for $J=2$ and $3$, and with no fundamental limit in going to higher rotational levels. Laser cooling and trapping of AlF is conceptually similar to the introduction of alkaline-earth atoms into cold atom physics, and is key to leveraging its spin-forbidden a$^3\Pi \leftarrow{}$X$^1\Sigma^+$ transition for precision spectroscopy and narrow-line cooling.
\end{abstract}

\maketitle

Remarkable progress has been made within the past decade in direct laser cooling and magneto-optical trapping of molecules. Since the first laser cooling \cite{Shuman2010}, radiation pressure slowing \cite{Barry2012} and magneto-optical trap (MOT) \cite{barry_magneto-optical_2014} of SrF, the same techniques have been applied to CaF \cite{truppe_molecules_2017, anderegg_radio_2017}, YO \cite{collopy_3d_2018}, and more recently, to BaF \cite{zeng_three-dimensional_2024}, CaOH \cite{vilas_magneto-optical_2022} and SrOH \cite{lasner_magneto-optical_2025}. Efforts to laser slow and trap CaH \cite{vazquez-carson_direct_2022}, CaD \cite{dai_laser_2024}, MgF \cite{pilgram_spectroscopy_2024}, YbF \cite{alauze_ultracold_2021}, AlCl \cite{daniel_hyperfine_2023} and CaOCH$_3$ \cite{Mitra2020} are now well underway. Although the phase-space density in red-detuned molecular MOTs is far lower than in their atomic counterparts, substantial improvement for molecules has been achieved by gray and $\Lambda$-enhanced molasses cooling \cite{truppe_molecules_2017,Ding2020}, blue-detuned MOTs \cite{burau_blue-detuned_2023,li_blue-detuned_2024} and narrow-line cooling \cite{mehling_narrowline_2025}. Following capture into a MOT, molecules have been transferred into conservative magnetic \cite{williams_magnetic_2018,McCarron2018} and optical \cite{anderegg_laser_2018, anderegg_optical_2019, Langin2021, Wu2021, jorapur_high_2024, vilas_optical_2024} traps, with state-of-the-art optical tweezer experiments realising deterministic entanglement between the rotational levels of individual molecules \cite{holland_-demand_2023, bao_dipolar_2023}. These successes complement the well-established and powerful method of associating ultracold atoms for molecular quantum science \cite{Ni2008, Moses2017, Bigagli2024}.

Thus far, \textit{only} species with $^2\Sigma^+$ electronic ground states have been successfully laser-slowed and captured into a MOT. This electronic structure results in molecules that are chemically reactive, difficult to produce efficiently \cite{wright_cryogenic_2023}, and are often unstable in bimolecular collisions \cite{Meyer2011,Sardar2023}. Moreover, for the $^2\Pi\leftarrow{} ^2\Sigma^+$ transitions used to laser-cool these molecules, simple optical cycling via a single rotational line is only possible for the first excited ($N=1$) rotational level of the ground state. Any other laser cooling scheme involves at least two rotational levels of the ground state, both coupled to the excited state by optical and/or microwave transitions, and is therefore more complex and less efficient. 

The situation is markedly different for molecules possessing $^1\Sigma^+$ electronic ground states. Many of these are both deeply bound and chemically stable, favoring efficient production in molecular sources. Further, for molecules with $^1\Pi\leftarrow{} ^1\Sigma^+$ optical transitions, all Q$(J)$ rotational lines are closed under electric dipole selection rules \cite{Fitch2021}. Provided the losses to excited vibrational levels are under control, optical cycling is then straightforward for \textit{any} rotational level with $N>0$. Finally, the separation of electronic states into spin-singlet and spin-triplet manifolds means narrow, spin-forbidden transitions are generally available in these systems, similar to those powerfully deployed in the alkaline-earth atoms. Until recently, detailed studies were restricted to BH \cite{Hendricks2014} and TlF \cite{Hunter2012}, because the most promising candidate molecules possess cooling transitions deep in the ultraviolet.

\begin{figure*}[tb]
    \includegraphics[width = \textwidth,trim={0 0.4cm 0 0}]{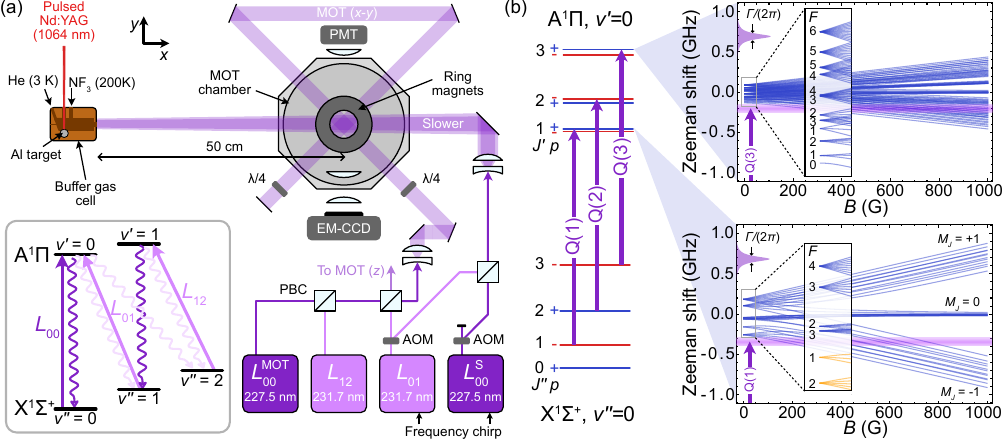}
    \caption{a) Schematic view of the experimental setup, showing the molecular beam source, magneto-optical trap and the laser configuration. The laser cooling scheme is shown as an inset on the left, where $L_{v'v''}$ labels the corresponding A$^1\Pi, v' \leftarrow{}$ X$^1\Sigma^+ , v''$ band excited in AlF. PBC = polarising beam splitter cube; AOM = acousto-optic modulator. b) Rotational and hyperfine structure of the lowest X$^1\Sigma^+,v''=0$ and A$^1\Pi, v'=0$ levels, showing the Q($J$) rotational lines excited by the $L_{00}$ lasers. Zeeman shifts are plotted on the right for the A$^1\Pi, v'=0$, positive parity levels with $J'=1$ and $J'=3$. Zoom-ins show the region $0<B<50$~G, and label the hyperfine levels by their total angular momentum quantum number $F$  \cite{Truppe2019}. Blue (orange) lines indicate levels with magnetic $g$-factor $g_F \geq 0$ ($< 0$). The natural linewidth of the transition, $\Gamma/(2\pi) = 84$~MHz, is indicated in the top left of each plot, with the $L_{00}^{\mathrm{MOT}}$ laser detuning for the MOT shown in the same color by shaded horizontal bars.}
    \label{fig:fig1}
\end{figure*}

Here, we demonstrate magneto-optical trapping of aluminum monofluoride (AlF). AlF is a collisionally stable diatomic molecule with a deeply bound (6.9~eV) $^1\Sigma^+$ electronic ground state; in particular, bimolecular collisions to form the metal difluoride are more than 1.0~eV endothermic for AlF. Our MOT operates on closed Q$(J)$ lines of the A$^1\Pi\leftarrow{}$X$^1\Sigma^+$ transition near 227.5~nm. We demonstrate a MOT for the first three excited rotational levels of the X$^1\Sigma^+$ state, and change rotational level in the trap by simply adjusting the laser frequencies used for slowing and trapping. The experiments reported here build on our prior studies of AlF’s electronic, vibrational, and hyperfine structure \cite{Truppe2019}, its suitability for direct laser cooling \cite{hofsass_optical_2021}, and production of a deep ultraviolet MOT of atomic Cd \cite{Padilla2025}. 

Figure \ref{fig:fig1}a shows a schematic view of our experimental setup and the laser cooling scheme. Pulses of cold AlF molecules are produced in a cryogenic buffer gas cell, are decelerated by frequency-chirped laser slowing after exiting the cell, and are loaded into the MOT 50~cm further downstream. Details of the molecular beam source are given elsewhere \cite{wright_cryogenic_2023}. Briefly, Al atoms are produced by laser ablation of a solid Al target and react with nitrogen trifluoride gas to produce AlF. The molecules collide with cold He buffer gas and leave the cell as a pulsed molecular beam, with a typical forward velocity of 150 m/s ($J=1$). The repetition rate of the molecular source is 1~Hz, and firing of the ablation laser defines $t=0$. The vacuum pressure in the MOT chamber is below $10^{-8}$~mbar.    

We use four deep ultraviolet (UV) laser systems, all based on continuous near-infrared Ti:Sa light, and reaching the deep UV via two successive stages of second harmonic generation (SHG). The notation $L_{v'v''}$ classifies the laser wavelength according to the A$^1\Pi,v'\leftarrow{}$X$^1\Sigma^+,v''$ band excited in AlF. Two systems operating near 227.5~nm provide the primary $L_{00}$ light for slowing ($L_{00}^S$, 200-250~mW) and trapping ($L_{00}^{\mathrm{MOT}}$, 200-300~mW). A further two systems, labelled $L_{01}$ (200~mW) and $L_{12}$ (80~mW) operate near 231.7 nm, and repump population leaking into $v''=1$ and $v''=2$. About $50\%$ of the total optical power is available at the MOT chamber due to losses along the chain of optics to the experiment.

To implement frequency-chirped laser slowing, we combine light from the $L_{00}^{\mathrm{S}}$ and $L_{01}$ lasers on a polarising beam splitter cube (PBC), and direct them into the experiment counter-propagating to the molecular beam. Each beam is rapidly shuttered using an acousto-optic modulator (AOM). Frequency chirping is implemented via the active cavity mirror of the Ti:Sa lasers, and a feed-forward system assists in maintaining SHG cavities at resonance. To load a MOT on the Q(1) line, both slowing lasers are linearly chirped by 0.63~GHz over 6–7~ms, corresponding to a velocity change of 140~m/s for molecules remaining in resonance with the $L_{00}^{\mathrm{S}}$ light.
\begin{figure*}[tb]
    \includegraphics[width = \textwidth,trim={0 0.4cm 0 0}]{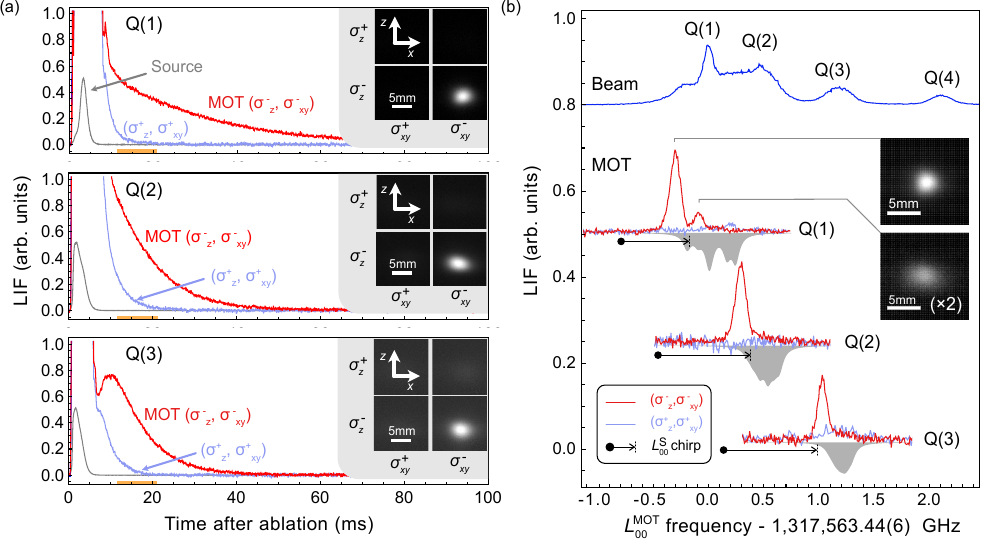}
    \caption{Magneto-optical trapping of AlF. a) Laser-induced fluorescence (LIF) traces demonstrating loading of the Q(1) (upper), Q(2) (middle) and Q(3) (lower) MOTs, for trapping (red) and anti-trapping (blue) configurations. Time-of-flight traces for the molecular beam arriving at the centre of the MOT chamber are shown in light grey. Insets: camera images taken using the four circular polarisation configurations of the $L_{00}^{\mathrm{MOT}}$ light; orange bars along the $x$-axis of the main panel show the camera exposure duration. (b) LIF spectra of the molecular beam (upper sub-panel) and each of the three MOTs (lower sub-panel) versus the $L_{00}^{\mathrm{MOT}}$ laser frequency. MOT spectra are vertically offset, and shown for trapping (red) and anti-trapping (blue) polarisation configurations. Black horizontal arrows show the  $L_{00}^\mathrm{S}$ frequency chirp used to load each MOT. Simulations (grey) pointing downwards show the Q$(J)$ lineshapes in zero magnetic field, including hyperfine structure of the A$^1\Pi,v'=0$ levels. Insets show camera images taken at the indicated laser frequencies.}
    \label{fig:fig2}
\end{figure*}

The MOT light is generated by combining $L_{00}^{\mathrm{MOT}}$, $L_{01}$ and $L_{12}$ light using two PBCs. One output of the second PBC provides trapping light in the $x$-$y$ plane, and the other output for the $z$-direction (not shown in figure \ref{fig:fig1}a). Quarter waveplates installed directly in front of the vacuum viewports generate circular polarised $\sigma_{z}^{\pm}$ ($\sigma_{xy}^{\pm}$) light for the $z$- ($x$-$y$) trapping beams of the MOT. Since the $L_{00}^{\mathrm{MOT}}$ and $L_{12}$ light is of opposite circular handedness to the $L_{01}$ light, we hereon refer only to the $L_{00}^{\mathrm{MOT}}$ polarisation, and call $(\sigma_z^{-},\sigma_{xy}^{-})$ the trapping configuration and $(\sigma_z^{+},\sigma_{xy}^{+})$ the anti-trapping configuration. A large magnetic field gradient is required for an effective MOT due to the large natural linewidth of the A$^1\Pi\leftarrow{}$X$^1\Sigma^+$ transition, $\Gamma/(2\pi) = 84$~MHz \cite{Truppe2019}. We use a pair of coaxially oriented NdFeB ring magnets mounted inside the MOT chamber, whose magnetisation axes are along $\pm z$, and whose inner surfaces are separated by $\Delta z = 1.95(5)$~cm. The resulting $z$-axis magnetic field gradient is 1.0~kG/cm near the MOT centre.

In figure \ref{fig:fig1}b, we show the structure of the lowest rovibrational levels of the X$^1\Sigma^+$ and A$^1\Pi$ states, with $J$ the total angular momentum quantum number excluding nuclear spin, and $p$ the parity. The nuclear spins of aluminum ($I_{\mathrm{Al}} = 5/2$) and fluorine ($I_{\mathrm{F}} = 1/2)$ result in a complex hyperfine structure and Zeeman effect. We focus here on the A$^1\Pi$ levels; in the X$^1\Sigma^+$ ground state the hyperfine and Zeeman interactions are one to two orders of magnitude smaller and unresolved. Zeeman shifts for all positive parity hyperfine levels with $J'=1$ (36 levels) and $J'=3$ (84 levels) are plotted in figure \ref{fig:fig1}(b) up to $1000$~G. Zoom-ins show the region $0<B<50$~G and label all hyperfine levels by their total angular momentum quantum number $F$, with $\mathbf{F}=\mathbf{J}+\mathbf{I}_{\mathrm{Al}}+\mathbf{I}_{\mathrm{F}}$. In the low-field region, the levels shift proportional to $g_F M_F B$, with $M_F$ the projection of the total angular momentum onto the magnetic field axis and $g_F$ a magnetic g-factor. A positive (negative) $g_F$ is indicated by blue (orange) curves in the zoom-ins of figure \ref{fig:fig1}b, and means that for a positive applied magnetic field, absorption of $\sigma^-(\sigma^+)$ polarised $L_{00}^{\mathrm{MOT}}$ light happens preferentially at red (blue) detuning. Molecules in the MOT sample the low and intermediate field regime, up to around 250~G; at sufficiently large fields, the levels separate according to the angular momentum projection quantum number $M_J'$. The complex level structure likely results in competition between confining and repulsive forces in the MOT \cite{Tarbutt2015}. Calculation of these forces is beyond the scope of this paper.

Figure \ref{fig:fig2}a shows a sequence of fluorescence traces, which demonstrate loading of a MOT with the $L_{00}^{\mathrm{MOT}}$ laser red-detuned from all Q$(1)$ hyperfine resonances. With the trapping polarisation configuration, molecules are loaded into the MOT from around $t=8$~ms, and fluorescence is observed for several tens of milliseconds (in the best case, beyond $t=100$~ms). The fluorescence trace with anti-trapping polarisations (blue) comprises only residual fluorescence from laser slowed molecules, and a free-flight arrival trace from the source is shown in light grey. We show camera images taken with each $L_{00}^{\mathrm{MOT}}$ polarisation configuration as an inset, observing a central cloud of molecules only with the trapping configuration. Together, these observations verify magneto-optical trapping. MOTs on the Q(2) and Q(3) lines are loaded by retuning the frequencies of the $L_{00}$, $L_{01}$, and $L_{12}$ light, and modifying the chirp parameters to account for the somewhat higher beam velocities. The corresponding fluorescence traces and camera images are shown for the Q(2) and Q(3) MOTs in the lower sub-panels of figure \ref{fig:fig2}a.

Fluorescence from the molecular beam and from the MOT is plotted versus the $L_{00}^{\mathrm{MOT}}$ laser frequency in figure \ref{fig:fig2}b. For the beam spectrum, we use a single, linearly polarised $z$-axis beam in the MOT chamber such that the fluorescence signal is insensitive to Doppler shifts. The MOT spectra are vertically offset according to the Q($J$) line, and in each case we compare the trapping (red) and anti-trapping (transparent blue) configurations. The frequency chirp applied to the $L_{00}^S$ laser is shown underneath each spectrum by a black arrow. Simulated Q($J$) lineshapes, including hyperfine structure in the A$^1\Pi$ levels, are shown pointing downwards in grey. For the Q(1) MOT, we observe two clear peaks. One of these occurs with the laser red-detuned from all excited hyperfine levels, and corresponds to a conventional red-detuned MOT. A second, smaller peak occurs with the $L_{00}^{\mathrm{MOT}}$ laser blue-detuned from the levels with negative $g_F$ (i.e., those levels colored orange in the zoom-in of figure \ref{fig:fig1}b), but red-detuned from all other levels. In the latter case, the cloud is more weakly confined and the fluorescence is about five times weaker, as is visible from camera images in the figure. For the Q(2) and Q(3) MOTs, we observe a single peak with the trapping light red-detuned from the beam resonance. The laser slowing and vibrational repumping make the trap highly selective on an individual rotational level, despite the close proximity of the Q$(J)$ lines.

\begin{figure}[tb]
    \includegraphics[width = \columnwidth,trim={0 0.4cm 0 0}]{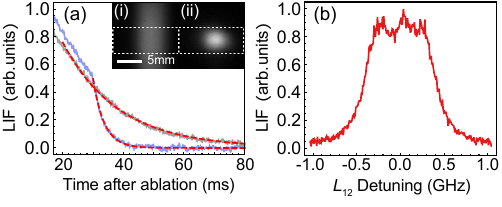}
    \caption{(a) Rapid switch-off of the $L_{12}$ light results in an increase in loss rate from the Q(1) MOT. Grey line: no switch-off of $L_{12}$. Transparent blue line: switch-off at $t=30$~ms. Red, dashed lines are fits to exponential decays from $t=30$~ms. Insets: camera images with identical imaging parameters for (i) Rayleigh scattering in nitrogen (1 bar) using a single $z$-axis $L_{00}^{\mathrm{MOT}}$ beam and (ii) Fluorescence from the Q(1) MOT. (b) Q(1) MOT fluorescence spectrum scanning the $L_{12}$ laser detuning.}
    \label{fig:fig3}
\end{figure}

To estimate the number of molecules captured into the Q(1) MOT, we use an additional AOM to switch off the $L_{12}$ light after loading the trap. The increase in loss rate from the trap, from $54$~s$^{-1}$ to $270$~s$^{-1}$, is shown in figure \ref{fig:fig3}a, and results from optical pumping to the $v''=2$ level of the X$^1\Sigma^+$ state. In a separate molecular beam setup, we measured the ratio of population pumped into the X$^1\Sigma^+,v''=1$ and $2$ levels following excitation with $L_{00}$ light to be 26(7). This means that the A$^1\Pi,v'=0 \rightarrow{}$X$^1\Sigma^+,v''=2$ branching ratio is about $1.8(5)\times 10^{-4}$, and that the average photon scattering rate in the Q(1) MOT is $1.2(3)\times10^6$~s$^{-1}$ per molecule ($\approx 2.3\times10^{-3}\Gamma$). The scattering rate is limited by the large two-level saturation intensity of 0.93~W/cm$^2$, and broad hyperfine structure of the Q(1) line; in the Q(2) and Q(3) MOTs we expect a slightly larger scattering rate, due to the decreasing span of the A$^1\Pi$ hyperfine structure. We observe that the photon counts on the camera are within a factor three of those from the Q(1) MOT. We calibrate the camera detection efficiency by filling the MOT chamber with 1 bar of $N_2$ gas and imaging Rayleigh-scattered laser light from a single $z$-axis trapping beam (27~mW), where the diameter of the beam is restricted to approximately match that of the MOT. Using the so-called `magic' angle of $\theta = 54.7 ^{\circ}$ between the (linear) polarisation of the laser light and the detector direction allows the isotropically averaged Rayleigh scattering cross section, $\sigma_{\mathrm{R}} = 1.9\times 10^{-25}$~cm$^2$ at $\lambda = 227.5$~nm, to be used \cite{Ityaksov2008}. The two insets to figure \ref{fig:fig3}a show camera images obtained using Rayleigh scattering, and for a typical Q(1) MOT. We compare their integrated intensities within the dashed rectangles (of length 5.2~mm along $z$), within which the total Rayleigh photon scattering rate is $7.2\times 10^{10}$~s$^{-1}$. From this, we arrive at an estimate of $6.4 \times 10^{4}$ molecules in the Q(1) MOT at the beginning of the camera exposure, and a peak density of $1.4\times 10^6$~cm$^{-3}$ at the MOT centre; for the Q(2) and Q(3) MOTs we estimate the number to be at least $10^4$ molecules. The uncertainty in these numbers is almost $50\%$, dominated by uncertainty in the branching ratio. Panel (b) shows the influence of scanning the detuning of the $L_{12}$ repump light on the Q(1) MOT fluorescence. The barely visible splitting of the A$^1\Pi,v'=1,J'=1$ excited level into three $M_J'$ components reflects the range of Zeeman shifts sampled by molecules across the MOT.   

The temperature in a red-detuned molecular MOT is in general significantly above the Doppler limit \cite{Devlin2016}, $\hbar \Gamma/(2k_B) = 2$~mK in our case. In combination with the relatively light mass, this makes it difficult to measure the temperature accurately via ballistic expansion within the magnetic field of the MOT. To provide an order of magnitude estimate, we switched off the Q(1) $L_{00}^{\mathrm{MOT}}$ light for between 0.5 and 3~ms before recapturing into the trap, and from this deduced a temperature of 16~mK. In a separate measurement, we observed damped harmonic oscillator motion in the trap following displacement with a short (250~$\mu$s) pulse of near-resonant $L_{00}^S$ light in the $x$-direction. The oscillation frequency, $\omega_x/(2\pi) = 175(6)$~Hz, in combination with the root-mean-square radius of the cloud of 1.4~mm, implies a temperature of 14~mK via the equipartition theorem.

The largest Q(1) MOT lifetime of around 27~ms was obtained with trapping beams enlarged to a ($1/e^2$) diameter of 13~mm and careful overlap of the two vibrational repump lasers. If we assume the observed loss rate is entirely the result of a leak in the optical cycle, then it must be at the level of few $10^{-5}$. This is far larger than loss due to the weak A$^1\Pi\rightarrow{}$a$^3\Pi$ decay ($10^{-6}-10^{-7}$) or from mixing of rotational levels due to the hyperfine interaction ($\lesssim 2\times 10^{-6}$ \cite{Truppe2019}); mixing of opposite parity levels in the A$^1\Pi$ state via the (quadratic) Stark effect would require stray electric fields of $2.5$~V/cm, which seems unlikely. Adding an additional $L_{23}$ vibrational repump laser to the Q(1) MOT light made no difference to the lifetime or trap number, from which we conclude that the loss probability to the $v''=3$ level of the ground state must be below the $10^{-5}$ level. For the Q(2) and Q(3) MOTs, the lifetime is shorter by about a factor two and four, respectively. We expect that a larger magnetic field gradient or yet larger trapping beams would be beneficial, to compensate for the weaker Zeeman effect in the A$^1\Pi$ state as $J'$ increases.  

Molecules in the MOT may be ionised from the A$^1\Pi$ state by absorbing a single photon from any of the trapping beams. Using a typical ionization cross-section of $10^{-17}$~cm$^2$ we calculate an upper limit for the ionization rate of $0.1$~s$^{-1}$ in our Q$(1)$ MOT so that in total, around one ion is generated per millisecond. Hence, the direct loss of molecules by ionisation has only a small effect on the observed trap lifetime. However, the electric field around an AlF$^+$ cation can potentially open a loss channel via the parity-mixing mechanism mentioned above. Whilst it seems unlikely to be limiting the lifetime of the MOT at present, the significance of this loss process will grow with the number and phase-space density in the MOT. 

In conclusion, we have demonstrated a deep ultraviolet magneto-optical trap for aluminum monofluoride, making AlF the first spin-singlet molecule to be directly laser-cooled and captured into a MOT. The MOT is effective for the three lowest Q($J$) rotational lines of the A$^1\Pi\leftarrow{}$X$^1\Sigma^+$ transition. Extension to higher rotational levels, using compact and inexpensive molecular beam sources, should be feasible. To fully exploit the spin forbidden a$^3\Pi \leftarrow{}$X$^1\Sigma^+$ transition of AlF, further cooling to 1~mK or below will be needed. Should this be successful, the benefits are multiplied, as this transition is (also) highly vibrationally diagonal \cite{walter_spectroscopic_2022-1}, and its three Q-branches are (also) rotationally closed. The only ultracold $^1\Sigma^+$ diatomic species realised as yet are the bialkali molecules, whose production relies on the availability of laser-cooled atoms amenable to photo- or magneto-association. These molecules are scarce in nature and weakly bound ($\sim1$~eV). By contrast, AlF is in many respects analogous to carbon monoxide, the most deeply bound diatomic molecule, and both $^{27}$AlF \cite{Ziurys1994} and the radioactive $^{26}$AlF isotopologue \cite{Kamiński2018} have been detected in space.

\section*{Acknowledgements}

\noindent We are grateful to Simon Hofs\"{a}ss and Sebastian Kray for their earlier work on the experiment, and to Lajos Palanki for helpful discussions. This project was financially funded in part by the European Research Council (ERC) under the European Union’s Horizon 2020 Research and Innovation Programme ``CoMoFun" (Grant Agreement No. 949119, S.T.), and from the European Commission through Project No. 101080164 ``UVQuanT” (S.T.).
\\

\section*{Data availability}

\noindent The data that support the findings of this article are openly available \cite{SupportingData}.
\\
\\
\noindent \textit{Author Contributions} - Conceptualization: S.T., G.M.; Analysis: B.S., S.W., J.E.P.C.; Investigation: S.W., J.E.P.C., J.C., P.K., P.A., R.T.;
Methodology: S.T., S.W., B.S.; Project administration: G.M., S.T.; Software: B.S., S.W.; Supervision:
S.T., G.M., S.W.; Validation: S.W., J.E.P.C., G.M., S.T., B.S.; Visualization: S.W., J.E.P.C., G.M.; Writing (original
draft): S.W., J.E.P.C.; Writing (review and editing): all authors.

%

\end{document}